\documentclass[prd,nofootinbib,tightenlines,showkeys,showpacs]{revtex4}
\usepackage{amsfonts}
\usepackage{amssymb}
\usepackage{amsmath}
\usepackage{indentfirst}
\usepackage[brazil]{babel}
\usepackage[latin1]{inputenc}

\usepackage{amssymb}
\usepackage{graphicx}
\usepackage{amsmath}
\usepackage{eurosym}
\usepackage{dsfont}
\usepackage{amsmath}
\usepackage{amssymb}
\usepackage{amsthm}
\usepackage{stmaryrd}
\usepackage{mathrsfs}
\usepackage{textcomp}
\usepackage{amsfonts}
\usepackage{txfonts}
\usepackage[T1]{fontenc}
\usepackage{color}

\begin{document}

\newcommand{\C}{\ensuremath{\mathbb C}}
\newcommand{\ds}{\displaystyle}
\newcommand{\eso}{\ensuremath{\mathfrak {so}}}
\newcommand{\gel}{\ensuremath{\mathfrak{gl}}}
\newcommand{\hM}{\ensuremath{\hat{M}}}
\newcommand{\mE}{\ensuremath{{\mathcal E}}}
\newcommand{\hMmn}{\ensuremath{\hat{M}_{\mu\nu}}}
\newcommand{\hx}{\ensuremath{\hat{x}}}
\newcommand{\ieso}{\ensuremath{\mathfrak {iso}}}
\newcommand{\igel}{\ensuremath{\mathfrak {igl}}}
\newcommand{\ka}{\ensuremath{\kappa}}
\newcommand{\Mmn}{\ensuremath{M_{\mu\nu}}}
\newcommand{\N}{\ensuremath{\mathbb N}}
\newcommand{\R}{\ensuremath{\mathbb R}}
\def\U{\mathcal{U}}
\newcommand{\Z}{\ensuremath{\mathbb Z}}


\def\warning#1{\begin{center}
\framebox{\parbox{0.8\textwidth}{\bf #1}}
\end{center}}
\def\look
{\framebox{$\clubsuit$}\,}

\renewcommand{\thesection}{\arabic{section}}
\renewcommand{\thesubsection}{\thesection.\arabic{subsection}}
\renewcommand{\theequation}{\arabic{equation}}
\newcommand{\pv}[1]{{-  \hspace {-4.0mm} #1}}

\def\A{\mathcal{A}}
\def\C{\mathbb{C}}
\def\N{\mathbb{N}}
\def\R{\mathbb{R}}
\def\Z{\mathbb{Z}}
\def\g{\mathfrak{g}}
\def\h{\mathfrak{h}}
\def\p{\partial}
\def\d{\delta}
\def\proof{\noindent \textbf{Proof. }}
\def\qed{$\blacksquare$}
\def\F{\mathcal{F}}
\def\G{\mathcal{G}}
\def\H{\mathcal{H}}
\def\K{\mathcal{K}}
\def\x{\hat x}

\def\BI{{\rm 1\!l}}


\begin{center}
{\Large \bf $\kappa$-deformation of phase space;
  generalized Poincar\'{e} algebras and $R$-matrix}
\end{center}
\bigskip

\begin{center}

\author[a]{S. Meljanac } {\footnote{e-mail: meljanac@irb.hr}}
\author[a]{A. Samsarov } {\footnote{e-mail: asamsarov@irb.hr}}  and
\author[b,a]{ R. \v{S}trajn} {\footnote{e-mail: r.strajn@jacobs-university.de}} \\ 

\bigskip
 
\affiliation[a]{Rudjer Bo\v{s}kovi\'{c} Institute,
 Bijeni\v{c}ka c.54, HR-10002 Zagreb, Croatia} \\ 
\affiliation[b]{Jacobs University Bremen, 28759 Bremen, Germany }

\end{center}
\setcounter{page}{1}
\bigskip

%

{\footnotesize{We deform a phase space ( Heisenberg algebra and corresponding
    coalgebra) by twist. We present undeformed and deformed tensor
    identities  that are crucial in our construction.
 Coalgebras for the generalized Poincar\'{e} algebras have been constructed.
 The exact universal $R$-matrix for the
   deformed Heisenberg (co)algebra is found. We show, up to the third
    order in the deformation parameter, that in the case of
$\kappa$-Poincar\'{e} Hopf algebra this $R$-matrix can be expressed in
    terms of Poincar\'{e} generators only. This implies that the
    states of any number of identical particles can be defined in a
     $\kappa$-covariant way.} \\ 
{\footnotesize{Keywords: $\kappa$-deformation, phase space,
    generalized Poincar\'{e} algebra }}}
 

\bigskip

\vfill{\hfill{\it{A.S. dedicates this work to the memory of his late brother Sa\v{s}a Safti\'{c}.}}}

\bigskip



\newpage

\vspace*{1cm}

\section{Introduction}

Quantum field theory (QFT) appeared as a result of the attempt to
construct a theory that describes the many-particle systems in accord
with the quantum mechanical principles and the principles of special relativity.
The Poincar\'{e} symmetry is thus incorporated in the theory from the
very beginning. It is believed that for particles with very high
energies the gravity effects become significant \cite{Dop1}, so that these
particles no more see the spacetime as smooth and continuous, but
instead they see it as quantized and fuzzy. The QFT constructed on
such deformed manifolds requires a new framework. Such framework is
provided by the noncommutative geometry, where the search for the
diffeomorphisms leaving spacetime invariant leads to deformation of 
Poincar\'{e} symmetry, with $\kappa$-deformed  Poincar\'{e} symmetry
being among the most extensively studied \cite{Luk1},\cite{MR},\cite{k-gn02}.
Much interest has been generated in the study of the physical
consequences emerging from the $\kappa$-deformation of Poincar\'{e} symmetry,
 e.g. construction of field theories \cite{klm00},\cite{klry09},\cite{ms11},\cite{mstw11},
electrodynamics \cite{hjm11},\cite{dj11} and geodesic equation \cite{Harikumar:2012zi} on $\kappa$-Minkowski spacetime
and a modification of the particle statistics \cite{kappaSt},\cite{gghmm08},\cite{Gupta:2011uz}. 
$\kappa$-deformed  Poincar\'{e} symmetry is algebraically described
by the $\kappa$-Poincar\'{e} Hopf algebra and is an example of
deformed relativistic symmetry that can possibly describe physical
reality at the Planck scale. The deformation parameter $\kappa$ is
usually interpreted as the Planck mass or the quantum gravity scale.

The deformation of the symmetry group can be realized through the
application of the Drinfeld twist on that group \cite{Chaichian:2004za}. It is argued 
that for the twisted (deformed) Poincar\'{e}  symmetry to be retained
in a quantum theory, it is necessary to implement twisted statistics \cite{Balachandran:2006pi},\cite{gghmm08},
with the form of the interaction also being dictated by the quantum
symmetry.
The virtue of the twist formulation is that the deformed (twisted)
symmetry algebra is the same as the original undeformed one and only
the coalgebra structure changes, leading to the same free field
structure as the corresponding commutative field theory.

In QFT one deals with the asymptotic (in and out ) scattering  states
that are quantum states of the free theory. Under the action of the Poincar\'{e}  symmetry
algebra, one particle asymptotic  states transform according to the
irreducible representation of the Poincar\'{e} algebra
$\mathfrak{iso}(1,3)$. To extend the action of the symmetry algebra from
one particle to two and many particle states, one needs a notion of
the coproduct. Hence, if $D$ is a representation of the symmetry
algebra $\mathfrak{G}$ (which is here $\kappa$-Poincar\'{e} algebra,
$\mathfrak{G} \equiv \mathfrak{iso}_{\kappa}(1,3) $ ), which acts on the space
$V$ of physical states (asymptotic in and out states) as $ | \phi \rangle
\longrightarrow D(\Lambda) | \phi \rangle $, with $\Lambda \in
\mathfrak{G} $ and $| \phi \rangle \in V, $  then this action of the
symmetry algebra can be extended to two particle states according to
\begin{equation}
 | \phi_1 \rangle \otimes | \phi_2 \rangle \longrightarrow (D \otimes
   D) \Delta (\Lambda) | \phi_1 \rangle \otimes | \phi_2 \rangle,
\end{equation}
where $\Delta : \mathfrak{G} \longrightarrow  \mathfrak{G} \otimes
\mathfrak{G} $ is the coproduct.
Since in noncommutative versions of QFT, in much the same way as in the
standard QFT, one is concerned with the states of many identical
particles, it is of importance to identify the exchange statistics
that particles obey upon deformation. It is known that the
(anti)symmetrization procedure in NCQFT, required to describe
deformed bosons and fermions, is carried out with the help of the
statistics flip operator $\tau$ (intertwiner), which needs to be
compatible with the deformation, i.e. with
the twisted symmetry algebra. This means that $[\Delta (\Lambda), \tau] = 0.$ 
The symmetry algebra that admits the existence of such flip operator
belongs to a class of bialgebras that possess a quasitriangular
structure, which means that a universal quantum $R$-matrix for such
symmetry algebras can be found.
We recall that $R$-matrix for the symmetry algebra $\mathfrak{G}$ is an invertible element $R \in
\mathfrak{G} \otimes  \mathfrak{G} $ having the property
\begin{equation}
 R \Delta (\Lambda) R^{-1} = \tilde{\Delta} (\Lambda),
\end{equation}
for all $\Lambda \in \mathfrak{G}. $ Here $\tilde{\Delta} (\Lambda)$
is the opposite coproduct of $\Delta (\Lambda),$ where the factors in
the tensor product are interchanged in every term.
 In this case the statistics flip
operator $\tau$ can be expressed as $\tau = \tau_0 R,$ where $\tau_0 $
is the undeformed flip operator, $\tau_0 (| \phi_1 \rangle \otimes |
\phi_2 \rangle) = | \phi_2 \rangle \otimes | \phi_1 \rangle,$ for any
$| \phi_1 \rangle, | \phi_2 \rangle \in V. $

 We believe that the further detailed study of
 the algebraic structure of $\kappa$-Poincar\'{e}, in particularly the
  study of its triangular quasibialgebra structure is of high importance.
The existence of the triangular quasibialgebra structure ensures that
there is a fully $\kappa$-covariant way to define states of many
identical particles (in any representation) in such a way that there
is a one to one correspondence between the states of
$\kappa$-deformed theory and the  states of undeformed theory.

There have been attempts in the literature to obtain 
$\kappa$-Poincar\'{e} Hopf algebra from the Drinfeld twist, but
neither of them so far succeeded to accomplish this
completely. Particularly, a full description of deformation of 
Poincar\'{e} algebra in terms of both, the twist and the $R$-matrix is missing.
The Abelian twists \cite{gghmm08},\cite{Bu:2006dm},\cite{klry08} and Jordanian
twists \cite{bp09},\cite{byk09} compatible with $\kappa$-Minkowski spacetime have been constructed.
 However, the problems with these twists were first that they are not
expressed in terms of the Poincar\'{e} generators only and the second, that
they give rise to a coalgebra that does not close within the
$\kappa$-Poincar\'{e} algebra, but instead runs out, generally into $\mathcal{U}(\igel(4)) \otimes \mathcal{U}(\igel(4))$.
On the other side, in Ref.\cite{Young:2008zm} an attempt was made to find
the $R$-matrix that lies within the Poincar\'{e} algebra. While the
authors succeeded to find the perturbative expansion of the $R$-matrix up to the
fifth order  in deformation parameter, they failed to construct its
exact form. Also, they
 missed to give the form
of the twist and the corresponding relation between the twist and the $R$-matrix.


In this paper we elucidate and solve the above problems in a different
 approach, using deformation of the Heisenberg algebra and coalgebra
 by twist. In Section 2 we introduce the undeformed
 Heisenberg algebra and construct the corresponding coalgebra. We also present a type of tensor
 exchange identities and show that the introduced coalgebra is
 compatible with them.
 These identities appear to be  crucial in our construction.
 In Section 3 we deform the Heisenberg algebra
 and coalgebra
by the twist deformation and get the deformed tensor exchange
 identities that are also compatible with the deformed coalgebra
 structure. This deformed Heisenberg algebra includes the $\kappa$-Minkowski spacetime.
By using the homomorphism of the coproduct and the introduced
 tensor exchange identities, in Section 4, continuing analysis from Ref.\cite{Kovacevic:2012an}, we give the coproducts for
 the  Poincar\'{e} generators, proposing two new methods of
 calculation, which are explicitly
 illusrated in few examples. In Section 5 we give the exact form of
 the universal $R$-matrix, generally for the deformed Heisenberg
 coalgebra and especially 
 for the $\kappa$-Poincar\'{e} Hopf algebra,
 and present the method by which it can be cast into a form including 
Poincar\'{e} generators only. Finally, in Section 6, we give a short conclusion.

\section{Heisenberg algebra and coalgebra}

Before going into deformed relativistic symmetries let us start with the undeformed Heisenberg algebra and the deformation
of a phase space (deformed Heisenberg algebra) including $\kappa$-Minkowski spacetime.
In the undeformed case Heisenberg algebra $\ H$ can be defined as an
algebra generated by 4 coordinates $x_{\mu}$ and 4  momenta $p_{\mu}$,
satisfying the following relations:
\begin{equation}
\left[ x_{\mu },x_{\nu }\right] =0, \qquad \left[ p_{\mu },x_{\nu } \right]
=-i\eta _{\mu \nu }\cdot 1, \qquad \left[ p_{\mu },p_{\nu }\right] =0,
\end{equation}
where $\mu, \nu = 0,1,2,3$ and
 $\eta_{\mu\nu}=(-,+,+,+)$ is diagonal metric tensor with Lorentzian
signature. Similarly, the symmetric algebra in $ \{ x_{\mu } \} $ is
 denoted by ${\mathcal{A}},$ and the symmetric algebra in $ \{ p_{\mu } \} $
is denoted by $T$. Using relation $p_{\mu }x_{\nu } - x_{\nu }p_{\mu }
 = -i\eta _{\mu \nu }\cdot 1, $ we can write $H ={\mathcal{A}}T.  $ \\

The action $h \triangleright f(x) $ for any $h \in H, ~ f(x) \in {\mathcal{A}},  $
is defined by $ x_{\mu } \triangleright f(x) = x_{\mu } f(x), ~
p_{\mu } \triangleright f(x) = -i \partial f(x) / \partial x^{\mu }   $
and by  the property:
\begin{equation} \label{compositionrule}
 h_1h_2 \triangleright f(x) =h_1\triangleright (h_2\triangleright f(x)),
\end{equation}
where $h_1,\,h_2 \in H$. Hence, $H \triangleright {\mathcal{A}} = {\mathcal{A}} $
is a $H$-module. The Leibniz rule $h \triangleright (f(x)g(x)),$ for
any $h \in H,$ is obtained by using Eq.(\ref{compositionrule}) and from $x_{\mu } \triangleright (f(x)g(x)) =
\alpha (x_{\mu } f(x))g(x) + (1-\alpha)f(x) (x_{\mu } g(x)) $ and
$p_{\mu } \triangleright (f(x)g(x)) =
 (p_{\mu } \triangleright f(x))g(x) + f(x) (p_{\mu } \triangleright g(x)), $
where $\alpha$ is an arbitrary real number. 

The guiding principle in this paper is the idea that the Heisenberg
algebra $H$ can
be endowed with a coalgebra structure. Namely, the Leibniz rule and the coproduct
$\Delta_0 $ are related by:
\begin{equation} \label{compositionrule1}
 h \triangleright (f(x)g(x)) = m_0 (\Delta_0 h) \triangleright (f(x)
 \otimes g(x) ),
\end{equation}
for every $h \in H$, and $m_0$ is the multiplication map.
From the Leibniz rule for $x_{\mu }$ it follows:
\begin{eqnarray}\label{copx}
 \Delta_0 x_{\mu} = \alpha x_{\mu} \otimes 1 + (1-\alpha ) 1 \otimes  x_{\mu}
    = \left[ x_{\mu } \otimes 1 \right] = \left[1 \otimes x_{\mu }
    \right] ~ \in {\mathcal{A}} \otimes {\mathcal{A}} / {\mathcal{R}}_0,
\end{eqnarray}
where $\left[ x_{\mu } \otimes 1 \right] = \left[1 \otimes x_{\mu } \right]$
is the equivalence class generated by the relations
${\mathcal{R}}_0
  \equiv x_{\mu} \otimes 1 - 1 \otimes  x_{\mu} =0. $
The relations ${\mathcal{R}}_0$ generate the equivalence classes on ${\mathcal{A}} \otimes {\mathcal{A}}$.
It can be shown that $\Delta_0 ({\mathcal{A}}) = {\mathcal{A}} \otimes {\mathcal{A}} / {\mathcal{R}}_0 $
is an algebra isomorphic to ${\mathcal{A}}$. Similarly, from the
Leibniz rule for $p_{\mu }$ it follows:
\begin{eqnarray}
    \Delta_0 p_{\mu} =p_{\mu} \otimes 1 + 1 \otimes p_{\mu}. \label{copp}
\end{eqnarray}
Note that $\Delta_0 (T) $ is an algebra isomorphic to $T$.

Generally, $\Delta_0 h $ for any $h \in H, $ can be obtained by 
the homomorphism $\Delta_0 (h_1h_2) = (\Delta_0 h_1) (\Delta_0 h_2) $
that follows from Eq.(\ref{compositionrule}). Note that
$\left[p_{\mu} \otimes 1, x_{\nu} \otimes 1 - 1 \otimes  x_{\nu} \right] \neq 0, $
 but $ \left[ \Delta_0 p_{\mu}, x_{\nu} \otimes 1 - 1 \otimes  x_{\nu}
  \right] = 0 $ and generally the consistency requirement
$ \left[ \Delta_0 h, x_{\nu} \otimes 1 - 1 \otimes  x_{\nu}
  \right] = \left[ \Delta_0 h, {\mathcal{R}}_0 \right] = 0,$
for all $h \in H, $ is satisfied.
Taking $h_1=p_\mu, \,h_2=x_\nu$, and using Eqs.\eqref{copx}, \eqref{copp} and $\mathcal{R}_0$, one gets
\begin{eqnarray}
\Delta_0(p_\mu x_\nu) &=& (\Delta_0p_\mu)(\Delta_0x_\nu) \nonumber \\
&=&  \left\{      \begin{array}{clrr}
& (p_\mu \otimes 1+1 \otimes p_\mu) (x_\nu \otimes 1) =p_\mu x_\nu \otimes 1+ x_\nu \otimes p_\mu =                          p_\mu x_\nu \otimes 1+1 \otimes x_\nu p_\mu \\
& (p_\mu \otimes 1+1 \otimes p_\mu) (1 \otimes x_\nu ) =p_\mu \otimes x_\nu +1\otimes p_\mu x_\nu = x_\nu p_\mu \otimes 1+1 \otimes p_\mu x_\nu
\end{array} \right .
\end{eqnarray}
In complete analogy, one also obtains four expressions for $\Delta_0(x_\nu p_\mu)$. It now follows, from these expressions, that
\begin{equation}
\Delta_0\Bigl( [p_\mu, x_\nu ]\Bigr) =\Delta_0(p_\mu x_\nu)- \Delta_0(x_\nu p_\mu) =[\Delta_0 p_\mu ,\Delta_0 x_\nu] =-i\eta_{\mu\nu} \cdot 1\otimes 1,
\end{equation}
showing that the coproduct $\Delta_0 h, \, h\in H,$ is compatible with
the relations $\mathcal{R}_0$.
Hence, $\Delta_0 (H) = \Delta_0 ({\mathcal{A}})  \Delta_0 (T) $ is the
algebra isomorphic to $H$.
The coalgebra on the undeformed Heisenberg algebra $H$ is also
undeformed. The coalgebra of the Heisenberg algebra $H$
 does not lead to a Hopf algebra. It is related to an algebraic
 construction whose mathematical structure is described in terms of a
 Hopf algebroid \cite{lu}. 
Note that the elements $L_{\mu \nu} =x_\mu p_\nu$ generate the
 $\mathfrak{gl}(4)$ algebra and satisfy $\Delta_0 L_{\mu \nu} = L_{\mu \nu} \otimes 1+1 \otimes L_{\mu \nu}$.

In the following, we interpret $\Delta_0 h$ as the operator
$ \Delta_0 h: \Delta_0 (H)
 \longrightarrow \Delta_0 (H), $ defined by
$ (\Delta_0 h) (\Delta_0 h_0) = \Delta_0 (h h_0) $ 
for all $h, h_0 \in H $.


\section{$\kappa$-deformation of Heisenberg algebra and coalgebra by twist}

We shall be interested in a deformation of the Heisenberg algebra $H$. We consider the family of twists given by 
\begin{equation}
  \mathcal{F} = \exp\left(i(\lambda x_k p_k\otimes A-(1-\lambda)A
    \otimes x_k p_k)\right),
  \label{twist}
\end{equation}
where $A = - a \cdot p,$ $\lambda $ is a real parameter 
and $a$ is a deformation fourvector, whose only non-zero component is
the time component $a_0.$ The deformation parameter $a_0$ can be used to model deformation at the
quantum gravity scale, in which case it corresponds to the inverse of
the Planck mass, $a_0 \sim \frac{1}{\kappa}. $
 The given family of twists represents a
subfamily of Abelian twists which satisfy all required properties,
including a cocycle and counit condition \cite{gghmm08}.
Now the deformed coalgebra structure can be obtained by using this
twist operator $\mathcal{F}$ to get
\begin{equation}
  \Delta h = \mathcal{F}\Delta_0 h \mathcal{F}^{-1}, 
  \label{defcops}
\end{equation}
for every $h$ in $H$.
By the same twist operation applied to  relations ${\mathcal{R}}_0,$ namely
 $ \mathcal{F} {\mathcal{R}}_0 \mathcal{F}^{-1} \equiv {\mathcal{R}}, $
 they transform into 
\begin{eqnarray}\label{idxi}
 x_i \otimes 1 & = & Z^{\lambda - 1} \otimes x_i Z^{-\lambda},\\
  x_0 \otimes 1 & = & 1 \otimes x_0 - a_0 ((1 - \lambda) \cdot 1 \otimes x_k p_k + \lambda x_k p_k \otimes 1),  \label{idx0}
\end{eqnarray}
where $Z=e^{A}$. Let us name the  relations (\ref{idxi}),(\ref{idx0}) obtained in this way by
${\mathcal{R}} $. Similarly as before, these relations induce the partition
of the algebra ${\mathcal{A}} \otimes {\mathcal{A}} $
 into equivalence classes, which together with the algebra $\Delta (T) $
 forms the algebra $ \Delta(H) = \Delta({\mathcal{A}})\Delta(T)  $. This algebra contains all 
 deformed coproducts $\Delta h, $ for all $h \in H. $
 Again $\Delta h $ can be understood as the
  operator acting on $ \Delta(H) $
 and acquiring values in the same algebra, namely
$ \Delta h :  \Delta(H) 
 \longrightarrow  \Delta(H) . $
In this case the twist operator is defined accordingly as 
$ \mathcal{F} :  \Delta_0(H)
 \longrightarrow  \Delta(H) $,
 and vice versa for the inverse twist $\mathcal{F}^{-1}$.

 
By twisting the undeformed coproducts according to Eq.(\ref{defcops}), one
finds the following deformed coproducts
\begin{eqnarray} \label{defcopxi}
  \Delta x_i & =& x_i \otimes Z^{\lambda} =Z^{\lambda-1}\otimes x_i, \\
  \Delta x_0 & =& x_0 \otimes 1 + a_0 (1 - \lambda)\cdot 1 \otimes x_k  p_k =1\otimes x_0 -a_0\lambda x_kp_k \otimes 1, \label{defcopx0} \\
   \Delta p_{i} &=& p_{i}\otimes Z^{-\lambda }+Z^{1- \lambda}\otimes p_{i},     \label{defcoppi} \\
 \Delta p_{0} &=& p_0\otimes1+1\otimes p_0. \label{defcopp0}
\end{eqnarray}

It is worth mentioning that due to the twist deformation being expressed in
terms of the similarity transformation, the homomorphic property of the undeformed
coproduct $\Delta_0 $ naturally extends onto the coproduct $\Delta, $
so that it is also a homomorphism,
\begin{equation} \label{homomorphism}
  \Delta (h_1 h_2) = (\Delta h_1) (\Delta h_2), \qquad  h_1, h_2 \in H. 
\end{equation}
In analogy to the concluding remark of section 2, choosing $h_1=p_\mu, \,h_2=x_\nu$ in \eqref{homomorphism}, together with relations $\mathcal{R}$, gives us four expressions for $\Delta(p_\mu x_\nu)$. Repeating the procedure for $\Delta(x_\nu p_\mu)$, we find that $\Delta\Bigl( [p_\mu, x_\nu ]\Bigr) =[\Delta p_\mu ,\Delta x_\nu]$, i.e. $\Delta h,\, h\in H$ is compatible with relations $\mathcal{R}$ ($\mathcal{R}\Delta h= \Delta h\mathcal{R}$).

The basic idea in this paper is that the coproduct for any element in $H$
can be calculated by using only two things. The first one is the homomorphism of the coproduct $\Delta $
and  the second one are the relations ${\mathcal{R}},$ expressed in
terms of the identities, Eqs.(\ref{idxi}) and (\ref{idx0}). 
Particularly, the coproducts for the Lorentz generators can be calculated
by using this method, which we show in the next section.
 This is made possible
 since according to the theory of realizations
\cite{ms06},\cite{starproductrealisations},\cite{km11}
 the Lorentz generators can be expressed in terms of
$x$ and $ p$ generators of the Heisenberg algebra.
To make the picture closed, let us mention that in the theory of
realizations, there is one realization corresponding to each twist element.
More concretely, having a twist
element $\mathcal{F},$ the corresponding realization  can be obtained as
\begin{equation} \label{twistrealization}
  \hat{x}_\mu = m_0 \bigg( \mathcal{F}^{-1} \triangleright (x_\mu \otimes id )  \bigg),
\end{equation}
where $m_0$ is the multiplication map.
 The family of twists \eqref{twist} leads to
\begin{equation} \label{hxl}
\hat{x}_i= x_i Z^{-\lambda}, \quad \hat{x}_0=x_0 -a_0(1-\lambda)x_kp_k.
\end{equation}
One can check that realizations \eqref{hxl} for $\hat{x}_\mu$ satisfy the commutation relations 
\begin{equation}
  \left[\hat{x}_\mu,\hat{x}_\nu\right] =i\left(a_\mu\hat{x}_\nu-a_\nu\hat{x}_\mu
  \right),
  \label{kappa}
\end{equation}
that are known to describe $\kappa$-Minkowski spacetime algebra.
In this way the $\kappa$-deformation of the Heisenberg algebra $H$ is made explicit.

Eq.\eqref{hxl}, along with \eqref{defcopxi}-\eqref{homomorphism}, can be used to derive $\Delta \hat{x}_\mu$
\begin{equation}
\Delta \hat{x}_i = \Delta(x_iZ^{-\lambda}) = (\Delta x_i) (\Delta Z^{-\lambda}) = \left\{ \begin{array}{clrr} & (x_i\otimes Z^{\lambda}) (Z^{-\lambda}\otimes Z^{-\lambda}) = x_iZ^{-\lambda}\otimes 1= \hat{x}_i\otimes 1 \\
& (Z^{\lambda-1}\otimes x_i) (Z^{-\lambda}\otimes Z^{-\lambda}) = Z^{-1}\otimes x_iZ^{-\lambda}= Z^{-1} \otimes \hat{x}_i,
\end{array} \right .
\end{equation}
and analogously for $\hat{x}_0$, with the result
\begin{equation}
\Delta \hat{x}_0= \hat{x}_0\otimes 1 = 1\otimes \hat{x}_0 -a_0p_kZ^{\lambda-1}\otimes \hat{x}_k .
\end{equation}
These results can be written compactly as
\begin{equation}
\Delta \hat{x}_\mu =\hat{x}_\mu \otimes 1= Z^{-1}\otimes \hat{x}_\mu -a_\mu p_\alpha^{L}\otimes \hat{x}^{\alpha},
\end{equation}
where $p_0^{L}=(1-Z^{-1})/a_0$ and $p_i^{L}=p_iZ^{\lambda-1}$. The calculated
 coproducts are in accordance with the Leibnitz rule for $\hat{x}_\mu$
 (see Eq.(31) in \cite{km11}).




\section{Generalized Poincar\'{e} algebras and coalgebras} \label{IV}

There are infinitely many ways of implementing (deformed) Poincar\'{e} algebras 
compatible with the $\kappa$-Minkowski spacetime \cite{Kovacevic:2012an}. If one writes an ansatz for the Lorentz generators in the form
\begin{eqnarray} \label{realizationmi0} 
\hat{M}_{i0} &=& x_{i}p_{0}F_{1}\left(A,b\right) -x_{0}p_{i}F_{2}\left( A,b\right) +a_{0}\left( x_{k}p_{k}\right) p_{i}F_{3}\left( A,b\right) +a_{0}x_{i}\vec{p}^{2}F_{4}\left( A,b\right) \\
\hat{M}_{ij} &=& M_{ij}=x_{i}p_{j}-x_{j}p_{i}, \label{realizationmij} 
\end{eqnarray}
the (deformed) commutation relations can be found in \cite{Kovacevic:2012an}.
The form of the Lorentz generators given by (\ref{realizationmi0})
 is compatible with the deformed Lorentz algebra (with
 $\kappa$-Poincar\'{e} algebra included), where the Minkowski
 metric is replaced with the more general metric, having generally
 nonlinear dependence on momenta. This is unlike the form of the
 metric dependence of momenta that emerged within the idea of
 relative locality \cite{AmelinoCamelia:2011bm},\cite{rel_loc}, where the metric dependence on momenta is linear.

Following the results of the previous section (relations
$\mathcal{R}$, $\Delta_0$ and $\Delta$ for $x_{\mu}$ and $p_{\mu}$),
we present two (equivalent) new methods for calculating the coproduct
of $\hat{M}_{i0}$. The first method includes the twist deformation
\begin{equation}
\Delta \hat{M}_{i0} = \mathcal{F}\Delta_{0}\hat{M}_{i0}\mathcal{F}^{-1} \label{dm1} 
\end{equation}
with the help of the twist, Eq.(\ref{twist}), while the second method
uses the realizations (\ref{realizationmi0}) and
(\ref{realizationmij}) and the homomorphism of the coproduct $\Delta$,
\begin{equation}
\Delta \hat{M}_{i0} = \Delta x_{i}\Delta p_{0}\Delta F_{1} -\Delta x_{0}\Delta p_{i}\Delta F_{2} +a_{0}\Delta x_{k}\Delta p_{k}\Delta p_{i}\Delta F_{3} +a_{0}\Delta x_{i}\Delta \vec{p}^{2}\Delta F_{4}. \label{dm2}
\end{equation}
In Eq.(\ref{dm1}) $\Delta_{0}\hat{M}_{i0}$ is not simply $\hat{M}_{i0}\otimes
1+1\otimes \hat{M}_{i0}$, but needs to be calculated by using
Eqs.\eqref{copx} and \eqref{copp}
 ($\Delta_0 \hat{M}_{i0} =\Delta_0x_i \Delta_0p_0 \Delta_0F_1 -...$), and
 $\Delta x, \, \Delta p$ are given by Eqs.\eqref{defcopxi}-\eqref{defcopp0}. The coproduct of $M_{ij}$, calculated in an analogous way, is primitive.

We consider three examples of the above\\
(i) $F_1=\frac{Z^{2-\lambda}-Z^{-\lambda}}{2A},\, F_2=Z^{\lambda} ,\,F_3=(1-\lambda)Z^{\lambda},\, F_4=-\frac{Z^{\lambda}}{2}$. \\
This is the case of the undeformed Lorentz algebra which, if extended
 by noncommutative coordinates, forms a Lie algebra \cite{Kovacevic:2012an}. For this case we
 use $M$ instead of $\hat{M}$. Using Eq.\eqref{dm1} or \eqref{dm2} and the relations $\mathcal{R}$, gives
\begin{equation}
\Delta M_{i0}=M_{i0}\otimes1+Z\otimes M_{i0}-a_0Z^{\lambda} p_j\otimes M_{ij}.
\end{equation}
(ii) $F_1=\frac{\sinh A}{A},\, F_2=1,\, F_3=F_4=0, \, \lambda=\frac{1}{2}$ \\
In this example (known as the standard basis in Refs.\cite{Luk1}) the Lorentz
algebra is deformed ($[\hat{M}_{i0},\hat{M}_{j0}] =-iM_{ij}\cosh
A$). $\Delta \hat{M}_{i0}$, calculated by \eqref{dm1} or \eqref{dm2}
and $\mathcal{R}$ (relations \eqref{idxi} and \eqref{idx0} for $\lambda=\frac{1}{2}$), is
\begin{equation}
 \Delta\hat{M}_{i0}= \hat{M}_{i0}\otimes Z^{-\frac12}+ Z^\frac12\otimes\hat{M}_{i0} +\frac{a_0}2\left(M_{ij}Z^\frac12\otimes p_j- p_j\otimes M_{ij}Z^{-\frac12}\right).
\end{equation}
(iii) $F_1=F_2=1, \, F_3=F_4=0$ \\
We denote the generators appropriate to this case by $\widetilde{M}$. They generate the undeformed Lorentz (if extended by $p_{\mu}$, Poincar\'{e}) algebra, different from the case (i). Eq.\eqref{dm1} or \eqref{dm2}, together with $\mathcal{R}$, gives
\begin{equation}
\Delta \widetilde{M}_{i0}= x_ip_0\otimes Z^{\lambda} +Z^{-(1-\lambda)}\otimes x_ip_0 -x_0p_i\otimes Z^{-\lambda} -Z^{1-\lambda}\otimes x_0p_i -(1-\lambda)a_0 p_i\otimes x_k p_kZ^{-\lambda} +\lambda a_0 x_k p_k Z^{1-\lambda}\otimes p_i.
\end{equation}
An illustration of the method is given by calculating the coproduct of $M_{ij}$ (which is the same in all three cases):
\begin{eqnarray}
\Delta M_{ij} &=& \Delta x_i\Delta p_j- \Delta x_j\Delta p_i \nonumber \\
&=& x_ip_j\otimes 1+ \underbrace{x_iZ^{1-\lambda}\otimes p_jZ^{\lambda}}_{1\otimes x_ip_j} -x_jp_i\otimes 1 -\underbrace{x_jZ^{1-\lambda}\otimes p_iZ^{\lambda}}_{1\otimes x_jp_i} \nonumber \\
&=& M_{ij}\otimes 1+1\otimes M_{ij}.
\end{eqnarray}
The same result is trivially obtained using the twist because $\Delta_0M_{ij}$ is primitive and $M_{ij}$ commutes with $A$ and $x_kp_k$.

Expressions for the coproducts obtained in (i), (ii) and (iii)
coincide with the known results in the literature. The cases (i) and (ii) are the examples of $\kappa$-Poincar\'{e} Hopf algebra, while in the case (iii) the algebra needs to be extended to $\mathfrak {igl}(4)$ in order to get a $\kappa$-deformed $\mathfrak{igl}$(4) Hopf algebra, consistent with $\kappa$-Minkowski spacetime.

\section{Flip operator and the universal R-matrix}

The ordinary flip operator $\tau_0 :\ H \otimes H \longrightarrow H \otimes H$
 is defined by $\tau_0 (h_1\otimes h_2)=h_2\otimes h_1,$ for any two
 elements $h_1, h_2$ in $H,$ and has the properties
\begin{equation}
 \tau_0^{2}=1\otimes 1,  \qquad \tau_0\,\Delta_0 h=\Delta_0 h\,\tau_0 ,
\end{equation}
for all elements $h$ in $H$.
We use it to define the twist $\widetilde{\mathcal{F}}$ and the coproduct $\tilde{\Delta}$
\begin{eqnarray}
& \widetilde{\mathcal{F}}: \Delta_0(H) \longrightarrow
  \tilde{\Delta} (H),
& \qquad  \widetilde{\mathcal{F}}=\tau_0\, \mathcal{F}\,\tau_0,  \\
 &\tilde{\Delta}h : \tilde{\Delta} (H)
 \longrightarrow \tilde{\Delta} (H),
& \qquad \tilde{\Delta}h= \tau_0\, \Delta h\, \tau_0=
\widetilde{\mathcal{F}}\, \Delta_0h \, \widetilde{\mathcal{F}}^{-1},
\end{eqnarray}
for every $h$ in $H$.
Here $\widetilde{\mathcal{R}}=\tau_0 \mathcal{R} \tau_0 
 = \widetilde{\mathcal{F}} {\mathcal{R}}_0 \widetilde{\mathcal{F}}^{-1} $ are the relations coming from the coproduct $\tilde{\Delta}$
\begin{eqnarray}
&& x_i\otimes 1=Z^{\lambda}\otimes x_iZ^{1-\lambda}, \label{rtxi} \\
&& x_0\otimes 1=1\otimes x_0 +a_0\lambda\cdot 1\otimes x_kp_k +a_0(1-\lambda) x_kp_k\otimes 1  \label{rtx0}
\end{eqnarray}
and $\tilde{\Delta} (H) = \tilde{\Delta} ({\mathcal{A}})
\tilde{\Delta} (T), $ where $\tilde{\Delta} ({\mathcal{A}})
={\mathcal{A}} \otimes {\mathcal{A}}/{\widetilde{\mathcal{R}}}.  $
The algebra $\tilde{\Delta} (H)$ contains all coproducts $\tilde{\Delta} h,$
for all $h$ in $H$. Note that which one among the relations ${\mathcal{R}}_0, {\mathcal{R}},
 \tilde{\mathcal{R}}$ needs to be applied in a given
 situation depends on the codomain of the operator in question.

The flip operator $\tau$ is defined with $\ \tau \, \Delta h=\Delta h \, \tau$. It follows that
\begin{equation}\label{tau}
\tau=\mathcal{F}\tau_0\mathcal{F}^{-1}= \tau_0\widetilde{\mathcal{F}}\mathcal{F}^{-1} =\tau_0R.
\end{equation}
The last equality in \eqref{tau} is the defining relation of the $R$-matrix
\begin{equation}
  R=\widetilde{\mathcal{F}}\mathcal{F}^{-1},  \qquad  R: \ \Delta(H) \longrightarrow \tilde{\Delta}(H).
\end{equation}
From the property $\tau^{2}=1\otimes 1$ and the definitions of $\tau$ and $R$, one gets
\begin{equation} \label{rdr}
R\,\Delta h\, R^{-1}=\tilde{\Delta}h,
\end{equation}
for every $h$ in $H$.
For the family of twists \eqref{twist}, the corresponding $R$-matrix is given by
\begin{equation}\label{r}
R=\exp{\bigl(i(A\otimes x_kp_k -x_kp_k\otimes A)\bigr)},
\end{equation}
and it satisfies
\begin{equation}
\tau_0 \,R\,\tau_0=R^{-1},  \quad R^{-1}: \  \tilde{\Delta}(H)
 \longrightarrow  \Delta(H).
\end{equation}
We note that if the $\star$-product is defined by $(f\star g)_{\mathcal{F}} =m_0\mathcal{F}^{-1}\triangleright (f\otimes g)_{\mathcal{R}}$, where $\mathcal{R}$ in the subscript denotes that $(f\otimes g)_{\mathcal{R}} \in \Delta ({\mathcal{A}})$, and we define $(f\star g)_{\widetilde{\mathcal{F}}} =m_0\widetilde{\mathcal{F}}^{-1}\triangleright (f\otimes g)_{\widetilde{\mathcal{R}}}$, then it follows that
\begin{equation} \label{sfstf}
(f\star g)_{\mathcal{F}}=(g\star f)_{\widetilde{\mathcal{F}}} \ \Rightarrow \ \mathcal{F}^{-1}\triangleright (f\otimes g)_{\mathcal{R}} = \widetilde{\mathcal{F}}^{-1}\triangleright (g\otimes f)_{\widetilde{\mathcal{R}}},
\end{equation}
where $(g\otimes f)_{\widetilde{\mathcal{R}}} \in \tilde{\Delta}({\mathcal{A}}). $
Multiplying the second equation in \eqref{sfstf} by $\widetilde{\mathcal{F}}$ from the left, one finds
\begin{equation}
R\triangleright (f\otimes g)_{\mathcal{R}}= (g\otimes f)_{\widetilde{\mathcal{R}}}.
\end{equation}

By direct calculation (inserting \eqref{r} and
\eqref{defcopxi}-\eqref{defcopp0} into the LHS of \eqref{rdr}), it is
easy to check that \eqref{r} is the exact $R$-matrix for the
generators of the Heisenberg algebra $H$
\begin{eqnarray}
&& R\Delta x_iR^{-1}= x_i\otimes Z^{\lambda -1} =Z^{\lambda}\otimes x_i =\tilde{\Delta}x_i \label{dtxi} \\
&& R\Delta x_0R^{-1}=x_0\otimes 1-a_0\lambda\cdot 1\otimes x_k p_k = 1\otimes x_0 +a_0(1-\lambda) x_kp_k\otimes 1= \tilde{\Delta}x_0 \label{dtx0} \\
&& R\Delta p_i R^{-1}=p_i\otimes Z^{1-\lambda}+Z^{-\lambda}\otimes p_i =\tilde{\Delta}p_i \\
&& R\Delta p_0 R^{-1}=p_0\otimes 1+1\otimes p_0 =\tilde{\Delta} p_0,\label{dtp_0}
\end{eqnarray}
and because of the homomorphism property, also for the entire algebra
$H$. The first equalities in
Eqs.\eqref{dtxi}-\eqref{dtp_0} were calculated by using $e^{\rho}he^{-\rho} =e^{ad\rho}h$, and in the second equality of Eqs.\eqref{dtxi} and \eqref{dtx0} relations $\widetilde{\mathcal{R}}$ (\eqref{rtxi}, \eqref{rtx0}) were used. \\

In \cite{Young:2008zm} the authors have used a perturbative approach
to construct the universal $R$-matrix for the $\kappa$-deformed
Poincar\'{e} algebra generated by $p_\mu$ and $\hat{M}_{\mu \nu}$ from
the standard basis of \cite{Luk1} (case (ii) in Section \ref{IV}) up
to terms of order $\kappa^{-5}$ ($a_0=-1/\kappa$). The $R$-matrix they
presented is expanded in terms of the deformed Poincar\'{e} generators
and possesses a wedge-product structure. They give no statement about
the existence or appearance of the matrix at higher orders of the
deformation parameter. We have shown here that the exact $R$-matrix is
precisely  given by \eqref{r}.

Our method for expanding $R=e^{\rho}, \, (\rho= i(A\otimes x_kp_k
-x_kp_k\otimes A))$ in terms of $p_\mu$ and $\hat{M}_{\mu\nu}$ is the
following. We wish to find $r_1,r_2, ...$ such that
$R=\exp(r_1+r_2+r_3+...)$, where $r_i$ is a series in $a_0,$
starting with $a_0^{i}$
 and  is linear in $\hat{M}_{\mu\nu}$ (we do not consider terms
 containing the momenta alone), for all $\, i=1,2,3,...$. It follows that $\rho +\frac{1}{2}\rho^{2}+\frac{1}{6}\rho^3+...=r_1+r_2+\frac{1}{2}r_1^{2} +\frac{1}{2}(r_1r_2+r_2r_1) +r_3+ \frac{1}{6}r_1^3+...$.

In the first order we get $r_1=\rho$. The next step is to write down
an ansatz for $r_1$, linear in $\hat{M}$ and $a_0$, with undetermined
coefficients : $-ia_0(c_1\hat{M}_{i0}\otimes p_i
+c_2\hat{M}_{i0}p_i\otimes 1+ d_1\cdot 1\otimes \hat{M}_{i0}p_i
+d_2p_i\otimes \hat{M}_{i0})$. In order to compare this expression
with $\rho$, in all the terms from $\rho$ and the ansatz, we move
$x_\mu$ to the right side of the tensor product using the relations
$\widetilde{\mathcal{R}}$ (relations \eqref{rtxi} and \eqref{rtx0} for
$\lambda=1/2$). We also use the fact that $\hat{M}_{i0}=x_ip_0-x_0p_i+
O(a_0^2)$ ($\hat{M}_{i0}= x_ip_0\frac{\sinh A}{A} -x_0p_i$, see
Section \ref{IV}). After equating the expressions, we get 7 equations
for the 4 coefficients ($c_1,\, c_2,\, d_1,\, d_2$). The system of
equations has a unique solution given by $c_1=-d_2=-1,\, c_2=d_1=0$,
so that the final result up to the first order in the deformation
parameter $a_0$ is
\begin{equation} \label{lukierskicritique}
r_1= -ia_0(p_k\otimes \hat{M}_{k0}-\hat{M}_{k0}\otimes p_k),
\end{equation}
namely, $r_1 = -ia_0(p_k\otimes \hat{M}_{k0}-\hat{M}_{k0}\otimes p_k)  = \rho + O(a_0^2). $
Using the relations to move $x_\mu$ generates terms of higher order in the deformation parameter, e.g., $\rho=-ia_0( p_kZ^{\frac{1}{2}}\otimes x_kp_0Z^{\frac{1}{2}} -p_0\otimes x_kp_k)= -ia_0\cdot 1\otimes x_k( p_k\otimes p_0 -p_0\otimes p_k) +O(a_0^2)$.

In the second order, $r_2$ is found to be equal to $\frac{1}{2} \rho^{2} -\frac{1}{2}r_1^{2}$ + terms of order $a_0^{2}$ coming from the use of relations $\widetilde{\mathcal{R}}$ ($\lambda=1/2$) in $\rho-r_1$. It follows that
\begin{equation}
r_2=0.
\end{equation}
The general ansatz linear in $\hat{M}$, of order $a_0^2$, has 10 coefficients. Equating it with 0 gives a system of 16 homogeneous equations. The equations have a unique solution given by all the constants equal to 0.

In the third order, $r_3=\frac{1}{6}\rho^{3} -\frac{1}{6}r_1^{3}$ +
 terms of order $a_0^{3}$ coming from the use of relations $\widetilde{\mathcal{R}}$ ($\lambda=1/2$)
 in $(\rho+ \frac{1}{2}\rho^{2} -r_1 -\frac{1}{2}r_1^{2})$ + terms of order $a_0^{3}$ coming from the expansion of $\hat{M}_{k0}$ in $r_1$. Summing up all the contributions, we find
\begin{eqnarray}
r_3 &=& -\frac{ia_0^3}{24}\cdot 1\otimes x_i \Bigl( 3p_0^3\otimes p_i -3p_i\otimes p_0^3 +3p_0\otimes p_ip_0^2 -3p_ip_0^2\otimes p_0 +2p_0^2\otimes p_ip_0 -2p_ip_0\otimes p_0^2 \nonumber \\
&& -2p_ip_jp_0\otimes p_j +2p_j\otimes p_ip_jp_0 -2p_ip_j\otimes p_jp_0 +2p_jp_0\otimes p_ip_j -2p_ip_0\otimes p_jp_j +2p_jp_j\otimes p_ip_0 \nonumber \\
&& -2p_0\otimes p_ip_jp_j +2p_ip_jp_j\otimes p_0 \Bigr).
\end{eqnarray}
As for the first and second order, the obtained expression is compared with the most general ansatz linear in $\hat{M}$ and of order $a_0^{3}$, which in this case contains 28 constants. We get a system of 35 equations for the 28 coefficients. Solving the equations leaves 3 coefficients undetermined, giving us a three-parameter family of solutions for $r_3$
\begin{eqnarray}
r_3 &=& -\frac{ia_0^3}{24} \Bigl( -\alpha_1 M_{ij}p_jp_0\otimes p_i +\beta_1p_i\otimes M_{ij}p_jp_0 -\alpha_2M_{ij}p_j\otimes p_ip_0 +\alpha_2 p_ip_0\otimes M_{ij}p_j \nonumber \\
&& -(\alpha_1+2)\hat{M}_{i0}p_ip_j\otimes p_j +(\beta_1+2)p_j\otimes \hat{M}_{i0}p_ip_j +\alpha_1\hat{M}_{i0}p_jp_j\otimes p_i -\beta_1p_i\otimes \hat{M}_{i0}p_jp_j \nonumber \\
&& -2\hat{M}_{i0}\otimes p_ip_jp_j +2p_ip_jp_j\otimes \hat{M}_{i0} +3\hat{M}_{i0}p_0^2\otimes p_i -3p_i\otimes \hat{M}_{i0}p_0^2 \nonumber \\
&& +2\hat{M}_{i0}p_0\otimes p_ip_0 -2p_ip_0\otimes \hat{M}_{i0}p_0 +3\hat{M}_{i0}\otimes p_ip_0^2 -3p_ip_0^2\otimes \hat{M}_{i0} \nonumber \\
&& -(\alpha_2+2)\hat{M}_{i0}p_i\otimes p_jp_j +(\alpha_2+2)p_jp_j\otimes \hat{M}_{i0}p_i \nonumber \\
&& +(\alpha_2+2)\hat{M}_{i0}p_j\otimes p_ip_j -(\alpha_2+2)p_ip_j\otimes \hat{M}_{i0}p_j \Bigr).
\end{eqnarray}
If we choose $\alpha_1=\beta_1$, we get a two-parameter family of solutions for $r_3$, which all have a wedge-product structure. $r_1, \,r_2$ and the special case of $r_3$ with $\alpha_1=\beta_1=-6, \, \alpha_2=-2$ coincide with the results found in \cite{Young:2008zm}.

Analogously, the $R$-matrix can be expanded in the generators $p_\mu$
and $M_{\mu \nu}$ (from the case (i) in Section \ref{IV}). However, an
easier way is to insert the known relation \cite{Kovacevic:2012an} between $M$ (for $\lambda=1/2$) and $\hat{M}$ ($\hat{M}_{i0}=M_{i0}Z^{-\frac{1}{2}}+\frac{a_0}{2}M_{ij}p_j, \, \hat{M}_{ij}=M_{ij}$) into the results presented above. 

Contrary to $M$, the generators $\widetilde{M}$, from the case (iii)
in Section \ref{IV}, cannot be related to $\hat{M}$ (see Ref.\cite{Kovacevic:2012an}). Consequently, the
universal $R$-matrix cannot be expanded in $\widetilde{M}$ and
$p$. One arrives at the same result when trying to repeat the
presented
 procedure with $\widetilde{M}$ instead of $\hat{M}$ (the system of equations for the coefficients of $r_3$ doesn't have a solution).
 The above conclusions can be easily extended for the case of arbitrary $\lambda$.

\section{Concluding remarks and discussion}

  In the following we clarify what from our perspective is a major
  physical motivation for studying the universal $R$-matrix and also try
  to justify a type of deformation carried out in the paper which
  keeps the track with the Hopf algebra formalism, resulting in a
  mathematical structure that can be shown to be related to a Hopf algebroid \cite{lu}.

Basically, the idea for applying a Hopf algebra formalism is grounded
in the fact that the
general relativity theory together with the uncertainty principle of
quantum mechanics leads to a class of models with spacetime
noncommutativity. In this setting the smooth spacetime geometry of
classical general relativity is replaced with a Hopf algebra at the
Planck scale. In our view, a particularly interesting example of Hopf
algebra is $\kappa$-Poincar\'{e} Hopf algebra, which provides an
algebraic setting for describing symmetry underlying the effective
NCFT that results from coupling quantum gravity to matter fields after
topological degrees of freedom of gravity are integrated out \cite{AmelinoCamelia:2003xp,Freidel:2003sp,Freidel:2005bb,Freidel:2005me}. 
 This particular type of Hopf algebra also appears in the context of
 Doubly Special Relativity theories \cite{AmelinoCamelia:2000mn,AmelinoCamelia:2000ge,Magueijo:2001cr}, which give one possible
 explanation for few puzzling experimental observations, such as
 GZK-paradox found in observations of UHECR's \cite{bird,takeda}  and multi TeV photons \cite{aharonian}
 coming from certain distant astronomical objects, as well as recently
 obtained astrophysical data originating from the GRB's \cite{abdo}.

An important questions arising from the noncommutative nature of
spacetime at the Planck scale are how does this noncommutativity affect the
very basic notions of physics, such as the particle statistics,
particularly the spin-statistics relation and how these changes can be
implemented into a  quantum field theory formalism
 to accommodate for these new features.
For example, it has been known for some time that quantum gravity can
admit exotic statistics \cite{sorkin,an1,an2}. That quantum gravity
 may lead to such a situation was first discussed in \cite{sorkin} in the context of quantum geons.

 In our view, the answer to the above question, or at least some
  glimpse of it, is
  given within the framework of quantum groups, i.e. Hopf algebras and
  can be expressed in terms of its  triangular quasibialgebra
  structure or universal $R$-matrix. To be more precise, given a
  scalar field $\phi$ and knowing $R$-matrix, the corresponding
  modified algebra of creation and annihilation operators (which
  captures the information about statistics) can be inferred via relation
\begin{equation} \label{b5}
\phi(x)\otimes\phi(y)- R \phi(y)\otimes\phi(x)=0.
\end{equation}
In other words, we are addressing the issue as to how would the spin-statistics
 relation of a usual boson (or fermion) look like at the Planck scale. 

In this paper we have introduced the undeformed Heisenberg algebra and constructed the corresponding coalgebra. 
 The coalgebra so constructed induces relations ${\mathcal{R}}_0$ and
it is shown that the coalgebra and ${\mathcal{R}}_0$ identity are compatible.
We deform the Heisenberg algebra and coalgebra by the twist belonging
 to an Abelian family of twists. This deformed algebra includes $\kappa$-Minkowski
 spacetime. It is shown that the coalgebra structure is compatible
 with the deformed tensor identities. The coproduct of any element in
 $H$ is obtained, particularly the coproducts for the Lorentz
 generators are found.
 We point out that these coproducts can also be
 obtained by applying  the same twist deformation that was used to
 deform the Heisenberg algebra.
    In this way we have shown that there  exists a twist deformation that
leads to $\kappa$-Poincar\'{e} Hopf algebra. This means that from this twist,
the coalgebra and quasitriagular ($R$-matrix) structures  can be obtained that
are expressed in terms of  Poincar\'{e} generators only.
 With $R$-matrix staying entirely within
$\kappa$-Poincar\'{e} algebra, we have ensured  that states of
any number of identical particles can be introduced in a
$\kappa$-covariant fashion.

 We have presented the exact form of
 the universal $R$-matrix for the Heisenberg (co)algebra $H$ in general
and for the $\kappa$-Poincar\'{e} Hopf algebra.
 We have also outlined the  method for finding the $R$-matrix in
 terms of Poincar\'{e} generators only, with calculations being
 carried out up to the third order in the deformation parameter.
We have found that in the 3rd order there is a three-parameter family
 of solutions. For the certain choice of the parameter values, there
 is a two-parameter family of solutions having a wedge structure.
For one special choice of the parameters, these results coincide with
 the results obtained in \cite{Young:2008zm}. 

The general form of $\kappa$-Poincar\'{e} and $\kappa$-deformed $\igel(4)$ Hopf algebras might
provide an appropriate setting for capturing the signals coming from  Planck scale and Quantum gravity effects.
Some of those signals could be found in modified dispersion
relations \cite{AmelinoCamelia:1997gz,Gambini:1998it,Alfaro:1999wd,AmelinoCamelia:2000zs},
 which can directly be related to the so-called photon time delay.
This phenomenon could be a result of $\kappa$-deformation \cite{bgmp10} and it could be connected with the similar effect
measured for high energy photons coming from the gamma ray bursts.
Other footprints of the Planck scale physics might include a change
in the algebra of creation and annihilation operators describing
particles moving near the horizon of the black hole \cite{Gupta:2011uz}.
The change
 in the oscillator algebra  leads to deformed statistics and it is directly related to the
deformation of the $R$-matrix \cite{gghmm08}.

\acknowledgments
We would like to thank D. Kova\v{c}evi\'{c} for useful
comments and discussion. One of us (S.M.) thanks J. Lukierski on the
critical remarks related to $R$-matrix and Eq.(\ref{lukierskicritique}). This work was supported by the Ministry of Science and
Technology of the Republic of Croatia under contract
No. 098-0000000-2865. R.S. thanks the DFG (GRK 1620) for funding while the work was finished.


\end{document}